\documentclass[english, aip, apl, reprint]{revtex4-2}
\usepackage{units}
\usepackage{color}
\usepackage{amstext}
\usepackage{amssymb}
\usepackage{amsmath}
\usepackage{textcomp}
\usepackage{graphicx}
\usepackage{epstopdf}
\usepackage[T2A]{fontenc}
\usepackage{upgreek}
\usepackage{hyperref}
\usepackage{orcidlink}

\begin{document}
\title{The role of small-angle electron-electron scattering in transverse magnetic focusing experiment}

\author{Dmitry~A.~Egorov\,\orcidlink{0000-0001-5668-7284}}
\email[Author to whom correspondence should be addressed:~]{d.egorov@g.nsu.ru}
\affiliation{Rzhanov Institute of Semiconductor Physics SB RAS, 13 Lavrentiev Ave., Novosibirsk, 630090, Russia}
\affiliation{Department of Physics, Novosibirsk State University, 2 Pirogov Str., Novosibirsk, 630090, Russia}

\author{Dmitriy~A.~Pokhabov\,\orcidlink{0000-0002-8747-0261}}
\affiliation{Rzhanov Institute of Semiconductor Physics SB RAS, 13 Lavrentiev Ave., Novosibirsk, 630090, Russia}
\affiliation{Department of Physics, Novosibirsk State University, 2 Pirogov Str., Novosibirsk, 630090, Russia}

\author{Evgeny~Yu.~Zhdanov\,\orcidlink{0000-0002-7173-6213}}
\affiliation{Rzhanov Institute of Semiconductor Physics SB RAS, 13 Lavrentiev Ave., Novosibirsk, 630090, Russia}
\affiliation{Department of Physics, Novosibirsk State University, 2 Pirogov Str., Novosibirsk, 630090, Russia}

\author{Andrey~A.~Shevyrin\,\orcidlink{0000-0003-0632-2636}}
\affiliation{Rzhanov Institute of Semiconductor Physics SB RAS, 13 Lavrentiev Ave., Novosibirsk, 630090, Russia}

\author{Askhat~K.~Bakarov\,\orcidlink{0000-0002-0572-9648}}
\affiliation{Rzhanov Institute of Semiconductor Physics SB RAS, 13 Lavrentiev Ave., Novosibirsk, 630090, Russia}
\affiliation{Department of Physics, Novosibirsk State University, 2 Pirogov Str., Novosibirsk, 630090, Russia}

\author{Alexander~A.~Shklyaev\,\orcidlink{0000-0001-7271-3921}}
\affiliation{Rzhanov Institute of Semiconductor Physics SB RAS, 13 Lavrentiev Ave., Novosibirsk, 630090, Russia}
\affiliation{Department of Physics, Novosibirsk State University, 2 Pirogov Str., Novosibirsk, 630090, Russia}

\author{Arthur~G.~Pogosov\,\orcidlink{0000-0001-5310-4231}}
\affiliation{Rzhanov Institute of Semiconductor Physics SB RAS, 13 Lavrentiev Ave., Novosibirsk, 630090, Russia}
\affiliation{Department of Physics, Novosibirsk State University, 2 Pirogov Str., Novosibirsk, 630090, Russia}

\date{11 June 2025}

\begin{abstract}
We demonstrate the crucial role of small-angle scattering in transverse magnetic focusing (TMF) in ballistic GaAs/AlGaAs heterostructures. Measurements in various samples show that the role significantly depends on their geometry. We propose a phenomenological model parameterizing this dependence with the angular acceptance of the detecting contact. This model is consistent with the diversity of experimental data and therefore enables accurate extraction of the key characteristic of inter-electron (e-e) interaction —-- the e-e scattering length —-- from TMF experiment, thus turning it into a uniquely effective tool for studying e-e scattering.
\end{abstract}

\maketitle

\begin{figure*}
\centering
\includegraphics[width=1\linewidth]{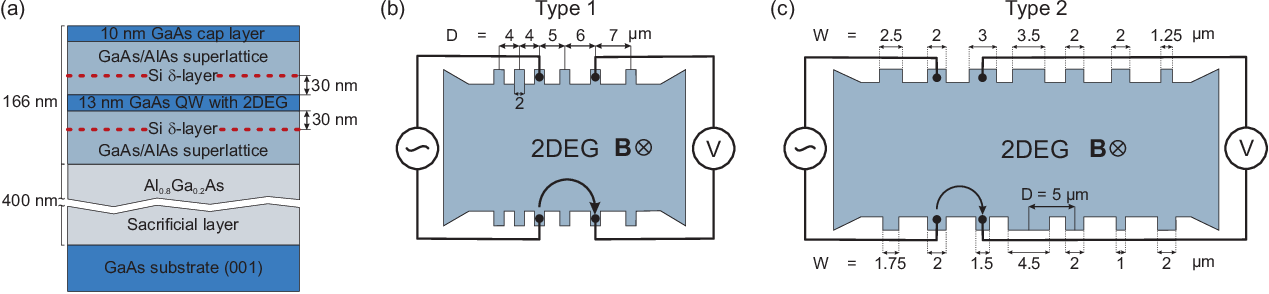}
\caption{\label{fig:Fig_1} 
(a) Heterostructure with a 2DEG. Non-local resistance measurement scheme and geometry of the Hall micro-bar devices with different distances between contacts (b, type 1) and widths of the contacts (c, type 2).}
\end{figure*}

The downscaling of semiconductor structures being in trend in their development, elevates electron-electron (e-e) interactions to a decisive role in certain phenomena. For instance, in quasi-one-dimensional systems, it can lead to Wigner crystallization, \cite{meyer2008} spatial splitting of a conductive channel into several ones,\cite{hew2009, Smith2009, ho2018, Pokhabov2019, Pokhabov2021, Pokhabov2020, sarypov2022} spontaneous symmetry breaking\cite{Thomas1996, Kristensen2000, Cronenwett2002, debray2009, Chen2008} and enhanced spin polarization \cite{pokhabov2018}. Somewhat counterintuitively, e-e interaction can enhance conductivity beyond the ballistic limit through a phenomena known as electron lubrication.\cite{gurzhi1962, guo2017} This effect suppresses momentum loss caused by impurities and rough walls, effectively facilitating more efficient electron transport. The diversity of e-e interaction effects requires careful study, well-known to remain challenging given the many-body physics involved. Even the fundamental characteristic of this interaction --- the e-e scattering length --- is difficult to both calculate and measure. Current theoretical approaches, including approximate models,are valid only at sufficiently low temperatures where $k_\mathrm{B}T \ll E_\mathrm{F}$,\cite{chaplik1971, giuliani1982} while many interesting interaction phenomena are observed in a much wider temperature range.\cite{keser2021, Sarypov2025} These include hydrodynamic effects \cite{dejong1995, gusev2018, krishnakumar2017, narozhny2019, ginzburg2021, Estradalvarez2025} associated with electron viscosity. Direct measurement of e-e scattering length is challenging as the scattering event conserves the total momentum of the electron subsystem, influencing the current indirectly (only through correlation effects, including modification of other scattering mechanisms).

A well-established, simple, and relatively direct method for experimental determining the e-e scattering length in two-dimensional electron systems is the transverse magnetic focusing (TMF) technique.\cite{tsoi1974, sharvin1965} The method exploits the sensitivity of non-local magnetoresistance, arising from peculiar resonant ballistic trajectories. Recent studies \cite{gupta2021Nature, Egorov2024} have successfully employed this approach to measure the temperature dependence of the e-e scattering length in GaAs-based 2DEGs, revealing general agreement with theoretical predictions \cite{giuliani1982} even beyond the model's formal applicability range with only the major flaw: thus measured e-e scattering length appears to be systematically lower than theoretical predictions. The origin of this discrepancy is understood to lie in the common interpretation of TMF results which does not fully take into account small-angle scattering effects, similar to those observed in impurity scattering length measurements.\cite{Spector1990} Elucidating the role of small-angle scattering in TMF experiments --- including its quantitative description --- could establish this technique as a reliable standard for determining the e-e scattering length, given its simplicity and accessibility.

In this work, we demonstrate that small-angle scattering significantly contributes to TMF experiments, and its proper accounting enables correct extraction of both the e-e scattering length and its temperature dependence. Using various ballistic samples based on GaAs/AlGaAs 2DEG, we show that sample geometry critically affects the extracted parameters. The experimental results are analyzed within an improved theoretical framework incorporating small-angle scattering.

The samples are fabricated from GaAs/AlGaAs heterostructures \cite{Pogosov2022, Pogosov2012} grown by molecular beam epitaxy (see Fig. \ref{fig:Fig_1}(a)). The structure consists of a 13 nm GaAs quantum well centered within a $166$~nm AlAs/GaAs short-period superlattice with two $\delta$-layers of doping silicon placed 30 nm from the quantum well. The superlattice contains low-mobility X-valley electrons near the $\delta$-layers, which do not conduct at low temperatures but effectively screen electrostatic impurity potential fluctuations, enhancing 2DEG mobility.\cite{Friedland1996} At 4.2 K, the electron density was $n = 7.8 \times 10^{11}$~cm$^{-2}$ with mobility $\upmu=2 \times 10^6$~cm$^2$/(V $\cdot$ s), corresponding to a Fermi energy of $28$~meV.

The 2DEG microstructures are created by a combination of photo- and electron lithography, with two geometries prepared. Type 1 samples are Hall-bars (see Fig. \ref{fig:Fig_1} (b)) in which all contacts have the same width of 2~$\upmu$m, but the distance $D$ between neighboring contacts varies from 4 to 7~$\upmu$m, and the width of the entire Hall-bar (i.e., the distance between opposite sides) is 25~$\upmu$m. Such contact arrangement allowed us to study TMF with cyclotron trajectory diameters ranging from 4 to 26~$\upmu$m. The type 1 samples are symmetric with respect to the longitudinal midline. Type 2 samples are Hall-bars (see Fig. \ref{fig:Fig_1} (c)), with fixed spacing between contact centers $D=5$~$\upmu$m but varying contact widths from 1.0 to 4.5~$\upmu$m. The width of the entire Hall-bar of type 2 is 15~$\upmu$m. The type 2 samples lack symmetry with respect to the longitudinal midline. All structures operate in the ballistic regime since their dimensions are smaller than the 30~$\upmu$m momentum relaxation length at the given electron density and mobility.

\begin{figure*}
\centering
\includegraphics[width=6.69in]{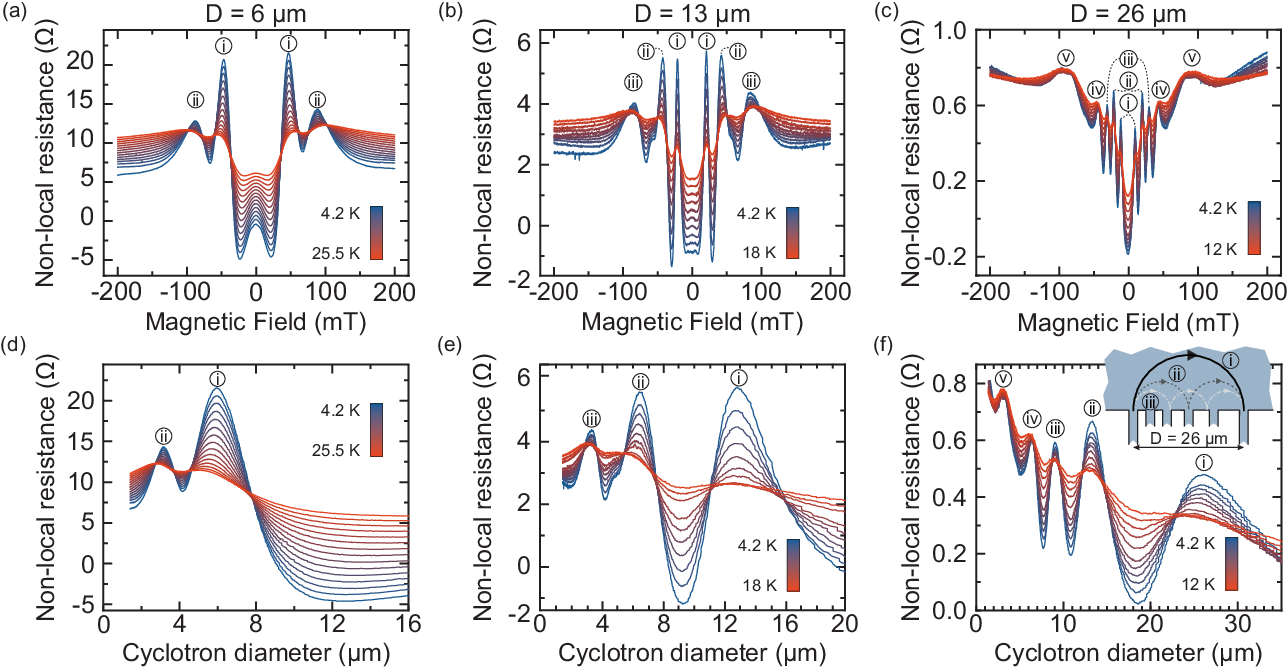}
\caption{\label{fig:Fig_2} 
Non-local resistance as a function of magnetic field at different temperatures with injector-detector distance (a) $D=6$~$\upmu$m (b) $D=13$~$\upmu$m, (c) $D=26$~$\upmu$m. Non-local resistance as a function of cyclotron diameter $d=\frac{{2\hbar \sqrt {2{\rm \pi }n} }}{{eB}}$ of electron trajectories with injector-detector distance (d) $D=6$~$\upmu$m (e) $D=13$~$\upmu$m, (f) $D=26$~$\upmu$m. The label "i" corresponds to TMF without reflection at the device boundary, "ii", "iii" etc. correspond to integer replicas with one, two, three etc. reflections, respectively (see inset in panel (f)).}
\end{figure*}

The samples are equipped with Au/Ni/Ge ohmic contacts. Measurements were performed in the linear response regime using lock-in detection techniques. Non-local resistance was measured: an alternating current with an amplitude of 1~$\upmu$A and a frequency of 23 Hz was passed through a pair of opposite contacts, one of which is considered as injector. A voltage was detected at the same frequency between another pair of opposite contacts, one of which, located on the same side as the injector, is considered as detector. The measurement scheme is shown in Fig. \ref{fig:Fig_1} (b-c). The non-local resistance was calculated as the ratio of the measured voltage and the passing current. During the experiment, magnetic field dependences of the non-local resistance were measured at fixed temperatures. Out-of-plane magnetic field varying from $-200$ to $200$~mT was applied. The temperature was varied from $4.2$ to $30$~K. To determine the actual values of electron concentration, standard measurements of the Hall resistance were performed in each individual thermocycle. The temperature suppression of the TMF magnetoresistance peaks was studied. The geometrical parameters were varied: for type 1 samples, the injector-detector distance was varied while their widths were fixed; for the type 2 samples, the width of the detector was varied while the width of the injector and injector-detector distance were fixed.

Fig. \ref{fig:Fig_2} (a-c) presents the magnetic field dependences of the non-local resistance at different temperatures, measured in the samples of type 1 using pairs of contacts separated by distances of $D = 6$, $13$, and $26$~$\upmu$m. It can be seen that the curves are symmetric with respect to magnetic field reversal, which directly results from the structural symmetry of the type 1 samples with respect to the midline along the Hall-bar. Each series of magnetic field dependences exhibits well-defined positive and negative non-local resistance peaks that are suppressed with increasing temperature.

\begin{figure*}
\centering
\includegraphics[width=6.69in]{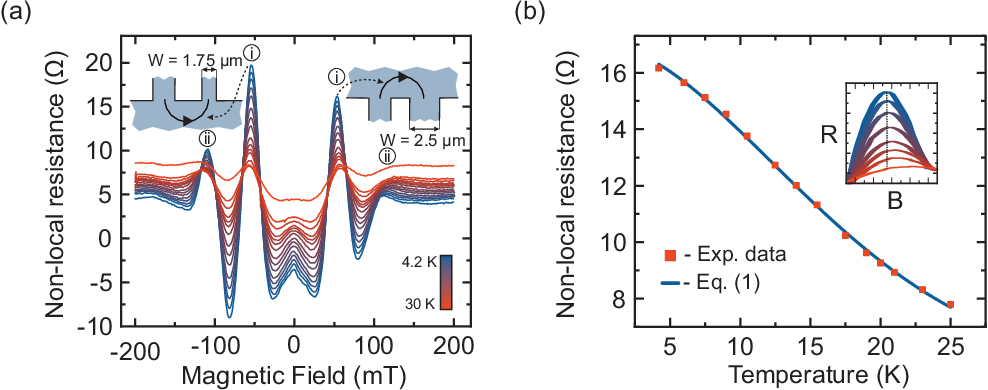}
\caption{\label{fig:Fig_3} 
(a) The non-local resistance as a function of magnetic field at different temperatures with injector-detector distance $D = 5$~$\upmu$m and detector width $W = 1.75$~$\upmu$m at $B<0$ (see left inset) and $W=2$~$\upmu$m at $B>0$ (see right inset). (b) The temperature dependence of the main focusing magnetoresistance peak amplitude with injector-detector distance $D = 5$~$\upmu$m and detector width $W = 2$~$\upmu$m.
}
\end{figure*}

Similar dependences were observed in type 2 samples with varying detector widths $W$, keeping the injector–detector distance $D$ fixed at $5$ and $10$~$\upmu$m (see Fig. \ref{fig:Fig_3} (a)).

The structural asymmetry between upper and lower contacts (see Fig. \ref{fig:Fig_1} (b)) enables the observation of magnetic focusing effects between both the upper (see the left inset in Fig. \ref{fig:Fig_3} (a)) and lower (see the right inset in Fig. \ref{fig:Fig_3} (a)) pairs of contacts in negative and positive magnetic fields respectively.

At certain temperatures we observe negative non-local resistance near zero magnetic field. This may originate from electron hydrodynamic effects \cite{Bandurin2016, gupta2021PRL}; discussion of their origin is beyond the scope of the present work. In the following analysis, we focus only at the ballistic features associated with TMF effects. 

The data presented in Fig. \ref{fig:Fig_2}(a-c) demonstrate the existence of several invariant points that all curves pass through regardless of temperature. The presence of such points indicates that the observed curve families are single-parametric. At sufficiently high temperatures (above 50 K) this behavior breaks down due to increased phonon scattering contribution. Experimentally, this is manifested as a shift of the curves away from these intersection points.

To relate the ballistic features to electron trajectories, we analyze the non-local resistance as a function of cyclotron orbit diameters $d_0=\frac{{2\hbar \sqrt {2{\rm \pi }n} }}{{eB}}$ (see Fig. \ref{fig:Fig_2} (d-f)). The dependences shown represent only partial segments of the corresponding graphs Fig. \ref{fig:Fig_2} (a-c) (positive magnetic fields). A distinct focusing peak (labeled "i") is observed at $d_0=D$, satisfying geometric resonance. It's finite width showing that system is tolerant to moderate deviations from $d_0=D$, and trajectories with $d_0$ slightly different from $D$ also contribute to the peaks. Such trajectories may connect, for example, either the closest or most distant parts of the injector and detector. However, at low temperatures, the focusing peak maxima occur precisely at the cyclotron diameter $d_0=D$. Among all possible cyclotron trajectories, these resonant orbits occupy the largest phase-space volume and thus yield the dominant contribution to the detected signal. The other peaks observed in the magnetoresistance curves are essentially integer replicas of the main focusing peak corresponding to electron motion with specular reflections from the device boundaries between the injector and detector.\cite{Heremans1999} These integer replicas appear at cyclotron orbit diameters equal to $d_0/2$, $d_0/3$ etc. For example, Fig. \ref{fig:Fig_2} (f) shows not only the primary peak at $d_0 = 26$~$\upmu$m ("i"), but also distinct replicas at $d_0 = 13$~$\upmu$m ("ii"), $d_0 = 8.7$~$\upmu$m ("iii"), and $d_0 = 6.5$~$\upmu$m ("iv"). We also note that the number of integer replicas increases with the injector-detector distance $D$. Interestingly, the appearance of these replica peaks is unaffected by the presence of other contacts between the injector and detector (see inset in Fig. \ref{fig:Fig_2} (f)). Experimental data show that electrons reflect from these contacts as from the sample boundaries. Notably, Fig. \ref{fig:Fig_3}(a) reveals weaker replica peaks for wide detectors (at $B=100$~mT) versus narrow ones (at $B=-100$~mT), attributed to reduced edge-to-edge distance degrading reflection conditions, causing both amplitude suppression and peak broadening.

We analyzed the temperature suppression of the main focusing peaks at the cyclotron diameter $d_0=D$, using a model assuming that non-local resistance is exponentially suppressed with a decrease in the scattering length $l_{\mathrm{sc}}$. \cite{gupta2021Nature, Egorov2024} In ballistic structures, electron-impurity scattering is negligible and essentially temperature-independent. Hence, the observed temperature suppression cannot be attributed to this mechanism and electron-impurity scattering may only affect pre-exponential factor $R_0$. As previously mentioned, electron-phonon scattering becomes significant only at high temperatures (above 50 K) which were not reached in this experiment. Thus, e-e scattering is considered as the dominant mechanism responsible for the temperature suppression of focusing peaks, meaning $l_{\mathrm{sc}} \approx l_{\mathrm{ee}}$. Consequently, the focusing peak amplitude depends on the e-e scattering length as follows: \cite{Egorov2024}
\begin{equation}\label{R_lee}
R = R_\infty + R_0 \exp{ \left[ -\upalpha\frac{l}{l_{\mathrm{ee}}} \right] },
\end{equation}
where $l$ is the trajectory length, $R_\infty$ and $R_\infty+R_0$ are the non-local resistance in the case of intense scattering ($l \gg l_{\mathrm{ee}}$) and weak scattering ($l \ll l_{\mathrm{ee}}$), respectively, $\upalpha$ is a dimensionless geometric factor. The e-e scattering length $l_{\mathrm{ee}}$ is set as:\cite{giuliani1982}
\begin{equation}\label{lee}
l_{\mathrm{ee}} ^ {-1} = \frac{(k_\mathrm{B} T)^2}{h E_\mathrm{F} v_\mathrm{F}} \left[ \ln{ \left( \frac{E_\mathrm{F}}{k_\mathrm{B} T} \right)} + \ln{\left(\frac{2 q_{\mathrm{TF}}}{k_\mathrm{F}} \right)} + 1 \right], \end{equation}
where $E_\mathrm{F}$ is the Fermi energy, $k_\mathrm{F}$ is the Fermi wavelength of the electron, $v_\mathrm{F}$ is the Fermi velocity, $k_\mathrm{B}$ is Boltzmann constant, $h$ is Planck constant, $q_{\mathrm{TF}}$ is the Thomas-Fermi wavenumber.

\begin{figure*}
\centering
\includegraphics[width=6.69in]{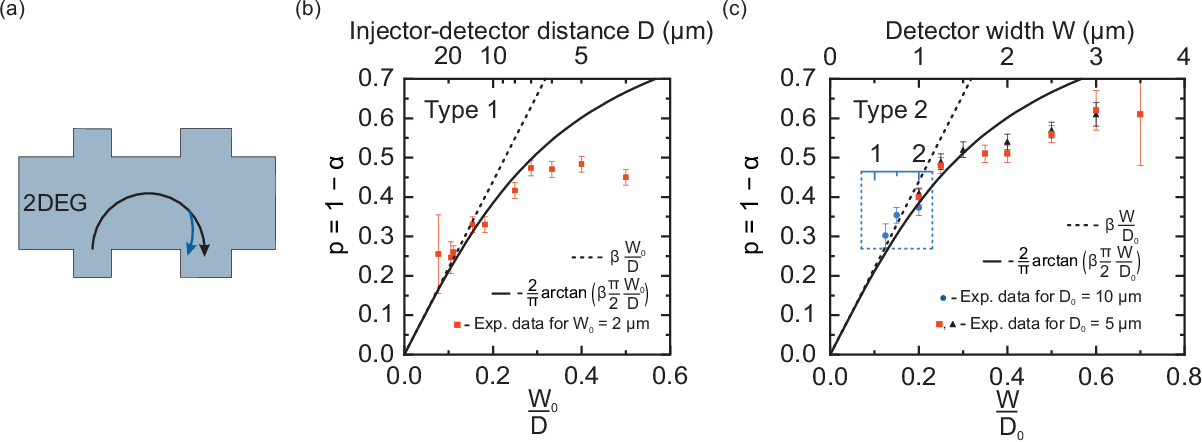}
\caption{\label{fig:Fig_4}
(a) The trajectory of an electron moving along a focusing trajectory in a Hall micro-bar, that reaches the detector despite scattering. (b) Dependence of the parameter $p=1-\upalpha$ on the ratio $W/D$ at a fixed detector width $W=W_0=2$~$\upmu$m. The top horizontal axis shows the injector-detector distance $D$. (c) Dependence of the parameter $p=1-\upalpha$ on the ratio $W/D$ at a fixed injector-detector distance $D=D_0$. The top horizontal axis displays the $W$-values for experimental data (red squares and black triangles two different samples of the same type) corresponding to $D=D_0=5$~$\upmu$m. The inset displays the $W$-values for experimental data (blue circles) corresponding to $D=D_0=10$~$\upmu$m.}
\end{figure*}

Thus, small-angle scattering weakens the role of e-e interactions in focusing peak suppression, consequently leading to an underestimated value of the e-e scattering length extracted from TMF experiments compared to its actual value in the 2DEG. This effect can be accounted for by introducing a numerical factor $\upalpha$ \cite{Spector1990} in Eq. \eqref{R_lee} representing

An electron scattered while following a resonant ballistic trajectory exhibits a bifurcation in its possible states. First, a long-angle scattering is likely to prevent the electron from reaching the detector, thereby reducing the peak height. Second, after a small-angle scattering, an electron can still be collected by the detector due to its finite width (see Fig. \ref{fig:Fig_4} (a)). Thus, the TMF experiment is to some extent tolerant to small-angle scattering. This tolerance should be properly taken into account to extract $l_\mathrm{ee}$ correctly, without systematic bias toward lower values. In the preceding analysis, this is naturally done via coefficient $\upalpha$ \cite{Spector1990} introduced in Eq. \eqref{R_lee}, which is in fact the probability of a scattered electron missing the detector. For further analysis, it is more convenient to use parameter $p = 1 - \upalpha$, which is the probability of a scattered electron reaching the detector. We assume that coefficient $\upalpha$ depends only on the TMF geometry and in what follows we focus at $p$ dependence on geometric parameters $W$ and $D$.

Fig. \ref{fig:Fig_3} (b) presents an example of the temperature dependence of the focusing peak amplitude. Since the peaks shift toward smaller diameters (less affected by e-e scattering) with increasing temperature (see Fig. \ref{fig:Fig_2} (d-f)), we specifically define the peak amplitude as the non-local resistance value at $d_0=D$. We fit the experimental data with the model function \eqref{R_lee}, using three fitting parameters $R_\infty$, $R_0$ and $\upalpha$. Fig. \ref{fig:Fig_3} (b) demonstrates that Eq. \eqref{R_lee} agrees well with the experimental data.

From the similar analysis done at type 1 samples (with different injector-detector distances $D$ ranging from $4$ to $26$~$\upmu$m) we extracted the dependence of parameter $p=1-\upalpha$ on $D$. Fig. \ref{fig:Fig_4} (b) presents the $p$ dependence on $\frac{W_0}{D}$ (the choice of the parameter $\frac{W_0}{D}$ is revealed further). The data demonstrate that $p$ increases with the decrease of the injector-detector distance $D$. Note that $p$ remains nearly constant and stabilizes at a value of about 0.45 for short trajectory lengths.

From an analogous analysis for type 2 samples (with different detector widths $W$ ranging from $1.0$ to $3.5$~$\upmu$m), we extracted the dependence of parameter $p=1-\upalpha$ on $W$. The corresponding dependence of $p=1-\upalpha$ on $\frac{W}{D_0}$ is presented in Fig. \ref{fig:Fig_4} (c). The data show that $p$ increases with the increase of the detector widths $W$.

To understand the observed behavior of coefficient $p$, we propose the following basic considerations regarding the dependence of the geometric factor $p$ on both the focusing trajectory length and detector width.

Let us consider several limiting cases. If the injector and detector were point-like (sufficiently narrow), small-angle scattering would be crucial since any scattering event would destroy the resonant trajectories and prevent electrons from reaching the detector. In other words, the focusing condition would become absolutely strict. Similarly, placing the injector and detector sufficiently far apart would impose the same zero-tolerance condition for scattering. In these limiting cases, the probability for an electron to reach the detector after any scattering vanishes implying $p=0$. Whereas, in actual experiments with contacts of finite widths and moderate injector-detector distances small-angle scattering effects must be accounted for.

An accurate calculation of small-angle scattering contribution to non-local resistance would be excessively complex, requiring contentious assumptions as well as complete consideration of all possible electron trajectories emitting from a finite-width injector. We assume that the coefficient $p$ is proportional to the probability of reaching the detector by a scattered electron that moved along a ballistic trajectory connecting the centers of the injector and detector. In the first approximation, in the case of small-angle scattering, this probability is linear with the detector's angular size, i.e., for smallest angles, proportional to $\frac{W}{D}$:
\begin{equation}\label{pLinear}
p=1-\alpha = \upbeta \frac{W}{D},
\end{equation}
here $\upbeta$ is a dimensionless universal parameter, which characterizes the averaged distance from the detector where scattering is most likely to occur. We note that Eq. \eqref{pLinear} satisfies the limiting cases described above.

From experimental $p$-dependencies at large injector-detector distances $D$ and small detector widths $W$, we extract the slope $\upbeta$ of the linear relation \eqref{pLinear}. In Fig. \ref{fig:Fig_4} (a) and (b), this straight line is shown as a dashed. The slope $\upbeta$ turns out to be the same and equal to 2.2 for all the samples used in both graphs \ref{fig:Fig_4} (a) and (b). However, Eq. \eqref{pLinear} describes the experimental data only at the smallest values of the $\frac{W}{D}$ ratio, indicating that a more general dependence on the detector's angular size should be used to describe larger values. We therefore propose the following generalized expression:
\begin{equation}\label{p}
p=1-\upalpha = \frac{2}{\rm\pi} \arctan \left( \frac{\rm \pi}{2} \upbeta \frac{W}{D} \right).
\end{equation}

The coefficient $p$ dependence on $\frac{W}{D}$ calculated using Eq. \eqref{p} with the same coefficient $\upbeta=2.2$ at fixed $W_0=2$~$\upmu$m, and at fixed $D_0 = 5$ and $10$~$\upmu$m is plotted in \ref{fig:Fig_4} (b) and (c), respectively. The described model of the dependence on the detector's angular size explains well the behavior of the coefficient $p$ under small-angle scattering conditions when $\upbeta \frac{\rm \pi}{2} \frac{W}{D} \lesssim 1$ (see Fig. \ref{fig:Fig_4} (b, c)), i.e., up to detector's angular sizes of 1 radian. Notably, the case of small-angle scattering corresponds to the narrowest focusing peaks, observed experimentally both at large $D$ (see Fig. \ref{fig:Fig_2} (c)) and at small $W$ (see Fig. \ref{fig:Fig_3} (a)).

In conclusion, we experimentally investigated the temperature suppression of non-local resistance focusing peaks, caused by e-e interaction. We established that this suppression is strongly affected by small-angle scattering, whose contribution we describe by the parameter $p=1-\upalpha$, characterizing the probability of a scattered electron reaching the detector. We propose a simplified model incorporating small-angle scattering that successfully predicts dependence of $p$ on experimental geometry. Moreover, this dependence can be fully characterized by a single parameter --- the detector's angular acceptance. We experimentally varied the angular acceptance injector-detector distance $D$, as well as with detector widths $W$. At small angular acceptance of the detector $\upbeta \frac{\pi}{2} \frac{W}{D} \lesssim 1$ the experimental results show good agreement with the model. In the case of large angular sizes of the detector $\upbeta \frac{\pi}{2} \frac{W}{D} \gtrsim 1$, it seems that the spatial and angular deviation of the contributing trajectories from the focusing trajectory has a strong influence. The explanation of such contribution requires a more detailed consideration, including the non-zero injector width and different electron departure angles during injection. It may also turn out that scattering at large angles cannot be described by introducing $\upalpha$ factor and requires a complex calculation of the probability of reaching the detector after scattering. We believe that the obtained results may be useful for the development of a technique for the experimental determination of the e-e scattering length in two-dimensional electron ballistic systems.

The work is supported by the Russian Science Foundation (grant No 22-12-00343-П -- experimental study) and the Ministry of Science and Higher Education of the Russian Federation (project FWGW-2025-0023 -- initial heterostructures characterization).

\section*{Author declarations}
\subsection*{Conflict of Interest}
The authors have no conflicts to disclose.

\subsection*{Author Contributions}
\textbf{Dmitry~A.~Egorov}: Investigation (equal); Formal analysis (equal); Visualization (equal); Data curation (equal); Writing - original draft (equal); Writing - review and editing (equal). \textbf{Dmitriy A. Pokhabov}: Conceptualization (equal); Supervision (equal); Investigation (equal); Formal analysis (equal); Validation (lead); Data curation (equal); Visualization (equal); Writing - original draft (equal); Writing - review and editing (equal). \textbf{Evgeny Yu. Zhdanov}: Investigation (supporting); Writing - review and editing (supporting). \textbf{Andrey~A.~Shevyrin}: Software (lead); Resources (equal); Writing - review and editing (equal). \textbf{Askhat K. Bakarov}: Resources (equal). \textbf{Alexander A. Shklyaev}: Resources (equal). \textbf{Arthur G. Pogosov}: Funding acquisition (lead); Supervision (equal); Conceptualization (equal); Writing - original draft (equal); Writing - review and editing (equal).

\section*{Data availability}
The data that support the findings of this study are available from the corresponding author upon reasonable request.

\section*{References}

\bibliography{Manuscript}

\end{document}